\documentstyle[12pt,amssymb,preprint,aps]{revtex}
\begin{document}
\baselineskip=.7cm

\begin{center}
{\Large {\bf {Singlet and triplet doped-hole configurations in  La$_{{\rm 2}}
$Cu$_{{\rm 0.5}}$Li$_{{\rm 0.5}}$O$_{{\rm 4}}$}}}
\end{center}

\vspace{1cm}

\begin{center}
{\bf V.I. Anisimov$^1$, S.Yu.~Ezhov$^1$\ and \ T.M. Rice$^2$}
\end{center}

\bigskip

\begin{center}
$^1$ Institute of Metal Physics, Russian Academy of Sciences, 620219
Ekaterinburg GSP-170, Russia\\$^2$ Theoretische Physik, Eidgen\"{o}ssische
Technische Hochschule-H\"{o}nggerberg, 8093 Z\"{u}rich, Switzerland
\end{center}

\vspace{5cm}

\noindent {\bf Abstract.} \ The ordered alloy La$_2$Li$_{0.5}$Cu$_{0.5}$O$_4$
is found to be a band insulator in local density approximation (LDA)
calculations with the unoccupied conduction band having predominantly $%
d_{x^2-y^2}$-symmetry and substantial weight in O $2p$-orbitals. This is
equivalent to a predominant local singlet configuration $d^9L$ or a low spin
Cu$^{3+}$-ion with both holes in orbits having $d_{x^2-y^2}$-symmetry, i.e.
Zhang-Rice singlets. A fairly modest reduction of the apical Cu-O bondlength
is sufficient to stabilize a high spin triplet Cu$^{3+}$-ionic configuration
with holes in both $d_{x^2-y^2}$ and $d_{3z^2-r^2}$ orbits in LDA + U
calculations. This leads us to identify the low energy triplet excitation
found in NQR studies by Yoshinari et al. as a local high spin Cu$^{3+}$%
-ionic configuration accompanied by a substantial reduction of the apical
Cu-O separation, i.e. a anti-Jahn-Teller triplet polaron.

\vfill\eject

\noindent
The substitution of lithium for copper in La$_2$CuO$_4$ is formally
equivalent to Sr doping on the La site as each donates one hole per dopant.
In lightly doped samples Li and Sr doping have essentially indistinguishable
effects on the magnetic properties and lattice structure~\cite{licuo1,licuo2}
. However in-plane substitutions for Cu and out-of-plane substitutions for
La were found to be very different in the conductivity. Li$^{+1}$ has ionic
radius essentially the same as that of Cu$^{+2}$, and brings a hole with it
into the plane. But the alloys La$_2$Cu$_{1-x}$Li$_{ix}$O$_4$ are never
metallic nor superconducting so that this hole must be localized in contrast
to the mobile holes introduced by out-of-plane substitutions. The solid
solubility of Li is such that complete filling of the copper band ( at 50\%\
Li) can be achieved. At this composition the Li and Cu ions form an ordered
superlattice~\cite{ref5} in which all Cu ions are surrounded by four
in-plane Li$^{+1}$ ions (1s$^2$, closed shell electronic configuration),
leading to isolated CuO$_4$ clusters. This compound was found to be a
diamagnetic insulator.

The Nuclear Quadrupole Resonance (NQR) study of La$_2$Cu$_{0.5}$Li$_{0.5}$O$%
_4$~\cite{fisk} reveals a magnetic excitation of the doped-hole state with
an energy of $\approx$ 130 meV. This energy is much smaller than the
estimation of the singlet-triplet splitting of the local Zhang-Rice singlet~%
\cite{zhang-rice} which is of the order of few eV. There are also
indications that this magnetic excitation is coupled to the charge
fluctuation or lattice distortion around the Cu site~\cite{fisk}.

La$_2$Cu$_{0.5}$Li$_{0.5}$O$_4$ is not the only compound with formally
trivalent copper which is a diamagnetic insulator. Another example is NaCuO$%
_2$~\cite{nacuo2}. Spectroscopy measurements and Configuration Interaction
(CI) calculations by Mizokawa et al.~\cite{fujimori} led them to conclude
that the ground state of NaCuO$_2$ is dominated by $d^9L$ ($L$: ligand hole)
configurations and not by $d^8$ as for a simple Cu$^{3+}$ ion. The ligand
holes $L$ are not metallic because strong $p-d$ hybridization between $d^8$
and $d^9L$ configurations leads to a split-off $d^9L$-like discrete state
above the oxygen continuum.

Calculation of the electronic structure of NaCuO$_2$ in Local (Spin) Density
Approximation (L(S)DA)~\cite{singh} showed that this compound can be
described as a conventional insulator with a band gap arising from simple
covalent effects ($p-d$ hybridization). The unoccupied band has more oxygen
character than copper thus confirming the conclusion from CI calculations
that "doped" holes are situated predominantly on oxygen orbitals. The key
structural element of the crystal structure of NaCuO$_2$ is the same as in
layered cuprates (for example La$_2$CuO$_4$): CuO$_4$ plaquettes with the Cu
atoms in the center of a square of oxygen atoms. However in layered
compounds the squares are corner-sharing and the Cu-O-Cu bond angles are
nearly 180$^\circ$ resulting in very broad $pd\sigma$-band, while in NaCuO$%
_2 $ the squares are edge-sharing and those angles are close to 90$^\circ$
and the corresponding bandwidth is much smaller.

In La$_2$Cu$_{0.5}$Li$_{0.5}$O$_4$ the CuO$_4$ plaquettes are separated by
Li ions, so that there is no Cu-O-Cu bonds at all and one can expect an even
narrower $pd\sigma$-band than in NaCuO$_2$. We have performed LDA
calculation for La$_2$Cu$_{0.5}$Li$_{0.5}$O$_4$ using the linearized
muffin-tin orbital method in an atomic-spheres approximation (LMTO-ASA)~\cite
{lmto}. The results (Fig.1) show that, indeed, the ground state is a
nonmagnetic insulator with a sizable gap value ($\approx$ 1 eV), compared to
the 0.3 eV value in NaCuO$_2$~\cite{singh}. The unoccupied band is again
rather narrow and with a symmetry of Cu centered $d_{x^2-y^2}$ orbitals. It
contains 40\% Cu3d orbitals and 60\% O2p states, indicating the strongly
covalent nature of the singlet ground state. The top of the valence band is
predominantly oxygen in origin, however there is significant admixture of $%
d_{3z^2-r^2}$ orbitals of copper.

Our results shows that in order to reproduce singlet ground state of the
doped hole in La$_2$Cu$_{0.5}$Li$_{0.5}$O$_4$ there is no need to take into
account Coulomb interaction corrections to the one-electron LDA. However the
magnetic excited state cannot be explored in a calculation scheme which does
not include Coulomb interactions inside the $d$-shell of copper. For example
the undoped cuprates experimentally are antiferromagnetic insulators while
LDA gives a paramagnetic metallic ground state. The reason for this
discrepancy is that while in LSDA the splitting between majority and
minority spin states is driven only by the exchange interaction with a value
of the Stoner parameter of $\approx$ 1 eV, the real driving force for the
antiferromagnetic insulator solution must be the much larger direct Coulomb
interaction parameter U $\approx$ 8e V. This contradiction was resolved in
the so-called LDA+U method where orbital-spin polarization caused by the
Coulomb interaction is directly taken into account~\cite{LSDA+U,lichtanis}.

The main idea of the LDA + U method is that LDA gives a good approximation
for the average Coulomb energy of $d$-$d$ interactions $E_{av}$ as a
function of the total number of $d$-electrons $N=\sum_{m\sigma} n_{m\sigma}$
where $n_{m\sigma}$ is the occupancy of a particular $d_{m\sigma}$-orbital:

\begin{equation}
E_{av}=\case1/2 U \;N(N-1) -\case1/ 4 J\;N(N-2).
\end{equation}
But LDA does not properly describe the full Coulomb and exchange
interactions between $d$-electrons in the same $d$-shell. So Anisimov {\it %
et al.}~\cite{LSDA+U,lichtanis} suggested to subtract $E_{av}$ from the LDA
total energy functional and to add orbital- and spin-dependent contributions
to obtain the exact (within a mean-field approximation) formula: 
\begin{eqnarray}
&&E =E_{LDA}-E_{av}\;+\;\case1/2\sum_{m,m^\prime,\sigma}U_{mm^\prime}
n_{m\sigma}n_{m^\prime -\sigma}  \nonumber \\
&&+\case1/2\sum_{m\neq m^\prime,m^\prime,\sigma}
(U_{mm^\prime}-J_{mm^\prime})n_{m\sigma}n_{m^\prime \sigma} \; .
\end{eqnarray}
Taking the derivative w.r.t. $n_{m\sigma}$ gives the orbital-dependent
one-electron potential: 
\begin{eqnarray}
&&V_{m\sigma}(\vec{r})=V_{LDA}(\vec{r})+
\sum_{m^\prime}(U_{mm^\prime}-U_{eff})n_{m^\prime -\sigma}  \nonumber \\
&&+\sum_{m^\prime \neq m}(U_{mm^\prime}-J_{mm^\prime} -U_{eff})n_{m\sigma} \\
&&+\;U_{eff}(\case1/2-n_{m\sigma})-\case1/4J \ .  \nonumber
\end{eqnarray}
with $U_{eff}=U-\case1/2J$. The Coulomb and exchange matrices $U_{mm^\prime}$
and $J_{mm^\prime}$ are expressed through the integrals over products of
three spherical harmonics and screened Coulomb and exchange parameters $U$
and $J$~\cite{LSDA+U}.

A nontrivial problem is what value of the screened Coulomb interaction $U$
to use. For insulators, such as late transition-metal oxides a good
approximation is to calculate static screening of the $d$-$d$ intrashell
Coulomb interaction in a supercell LDA calculation~\cite{Anisimov10}.

The question is what symmetry should the lowest energy excited states have?
The band gap separates states which are both mainly oxygen but due to the
hybridization with Cu3d orbitals the states have symmetry of $x^2-y^2$ and $%
3z^2-r^2$ for the unoccupied and occupied bands respectively. The $x^2-y^2$
band is higher in energy due to the Jahn-Teller-distorted CuO$_6$ octahedra
in La$_2$Cu$_{0.5}$Li$_{0.5}$O$_4$ crystal structure: the length of the Cu-O
bond in the ab-plane is 1.8 \AA \ \ but the distance to the apical oxygen is
larger - 2.4 \AA~\cite{ref5}. It follows that the lowest energy excitation
will be from the (formally) $d_{x^2-y^2\uparrow}d_{x^2-y^2\downarrow}$
singlet configuration to the $d_{x^2-y^2\uparrow}d_{3z^2-r^2\uparrow}$
triplet configuration.

We have performed LDA+U method calculations for both singlet and triplet
configurations. We find that a starting triplet configuration is not a
stable solution but self-consistently converges to the singlet solution.
(The singlet solution of LDA+U method is practically the same as for pure
LDA, because in the absence of the orbital-spin polarization the LDA+U
correction to LDA is irrelevant.) However, if the copper-apical oxygen bond
length is contracted by 16\% (0.38\AA ), then a stable {\it triplet solution
appears and becomes the ground state}. For 16\% contraction the total energy
of the magnetic solution is still 0.06~eV higher than the nonmagnetic one,
however already for 17\% contraction the triplet energy is 0.1~eV lower.

It is instructive to follow the changes, with copper-apical oxygen bond
contraction, of the two bands below and above Fermi energy (Fig.2). For the
undistorted structure (Fig.2a) those bands (the lower one of $d_{3z^2-r^2}$
symmetry and the higher one of $d_{x^2-y^2}$) are well separated from each
other and the admixture of the $d_{3z^2-r^2}$-orbitals to the valence band
is relatively small. With distortion the hybridization of the Cu3$%
d_{3z^2-r^2}$-orbitals with the apical oxygen 2p$_z$ orbitals becomes
stronger and the antibonding band goes higher in energy and the admixture of
the $d_{3z^2-r^2}$ -orbitals in this band becomes stronger (Fig.2b). In
order for the magnetic state to be stable, the splitting between spin-up and
spin-down bands must become large enough that the top of the spin-up $%
d_{x^2-y^2}$ band does not overlap with the bottom of spin-down $d_{3z^2-r^2}
$ band (in other words magnetic energy must overcome kinetic energy). The
Fig.2c shows that for the critical value of the distortion (16\%) this
condition is nearly satisfied.

Similar effects were found in a supercell LDA+U calculation for a doped hole
in La$_{2-x}$Sr$_x$CuO$_4$~\cite{polaron}. In that work two solutions were
found for a hole introduced in the CuO$_2$ plane: one solution had the
symmetry $x^2-y^2$ with the hole spin antiparallel to the d-hole of Cu atom,
while the other one had $3z^2-r^2$ symmetry and spin parallel to the Cu
spin. The latter solution was present only for a contracted apical Cu-O
distance (0.26\AA contraction). The total energy calculation in a
full-potential scheme including the lattice relaxation showed that the total
energy minima for these two solutions were very close in energy with the
triplet state only 54 meV higher than the ground state singlet.

As the hopping between CuO$_6$ octahedra in La$_2$Cu$_{0.5}$Li$_{0.5}$O$_4$
is smaller than in La$_2$CuO$_4$, one would expect more localized states in
the former and, hence, a larger separation between the singlet ground state
and excited triplet state. The value found in NQR measurements~\cite{fisk}
130 meV looks quite reasonable from this point of view.

We would like to emphasize that while the excited magnetic state is a
triplet with $S$=1, the actual magnetic moment residing in Cu$3d$ orbitals
found in LDA+U calculation is very small: 0.8 $\mu_B$. That is due to the
fact that the bands (or Wannier orbitals) of $x^2-y^2$ and $3z^2-r^2$
symmetry have only 40\% of Cu3d orbitals contribution, being mainly oxygen
in origin as implied by the predominance of $d^9L$ many electron
configurations in model CI calculations~\cite{fujimori}.

In conclusion, we find that LDA correctly gives a band insulator ground
state for La$_2$Li$_{0.5}$Cu$_{0.5}$O$_4$ with the unoccupied conduction
band having predominantly $d_{x^2-y^2}$-symmetry and substantial weight in O 
$2p$-orbitals -- a state equivalent to a formal valence Cu$^{3+}$ in a low
spin configuration or a Zhang-Rice singlet. A reduction of the Cu-O apical
distance stabilizes a high spin configuration and this leads us to identify
the 130 meV triplet excitation observed in NQR with such a local
configuration i.e. an anti-Jahn-Teller polaron. Two aspects of the
experiments remain to be clarified -- the origin of the low energy nuclear
spin relaxation process which dominates at lower temperature at $T\lesssim $
170K and secondly the absence of a significant activated contribution to the
uniform susceptibility from the 130 meV triplet excitations -- at least
below room temperature.

\bigskip

The work was partly supported by Russian Basic Research Foundation (RFBR
grant 96-02-16167). One of us (VIA) wishes to thank the `Zentrum f\"ur
Theoretische Studien' at the Institute for Theoretical Physics, ETH-Z\"urich
for hospitality.

\begin{figure}[tbp]
\caption{The total (a) and partial (O2p (b) and Cu 3d (c)) densities of
states (DOS) for La$_2$Li$_{0.5}$Cu$_{0.5}$O$_4$ calculated in the
undistorted crystal structure. The bands formed by different d-orbitals are
marked by arrows.}
\label{fig.1}
\end{figure}
\begin{figure}[tbp]
\caption{The partial Cu 3d densities of states (DOS) for La$_2$Li$_{0.5}$Cu$%
_{0.5}$O$_4$. Upper panel (a): nonmagnetic solution in the undistorted
crystal structure. Middle panel (b): nonmagnetic solution with 16\%
contraction of the copper-apical oxygen bond length. Lower panel (c):
magnetic solution with 16\% contraction of the copper-apical oxygen bond
length.}
\label{fig.2}
\end{figure}

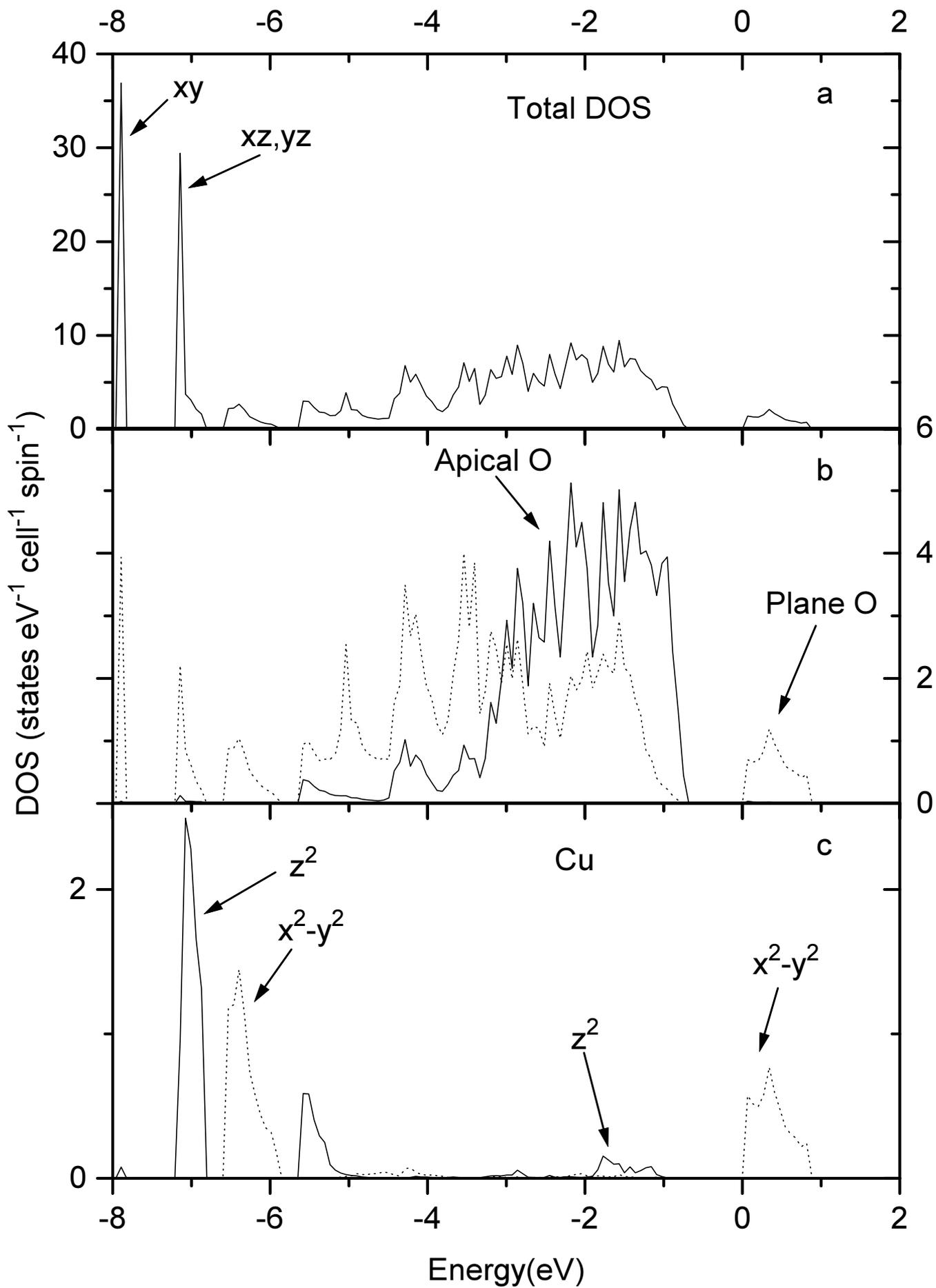
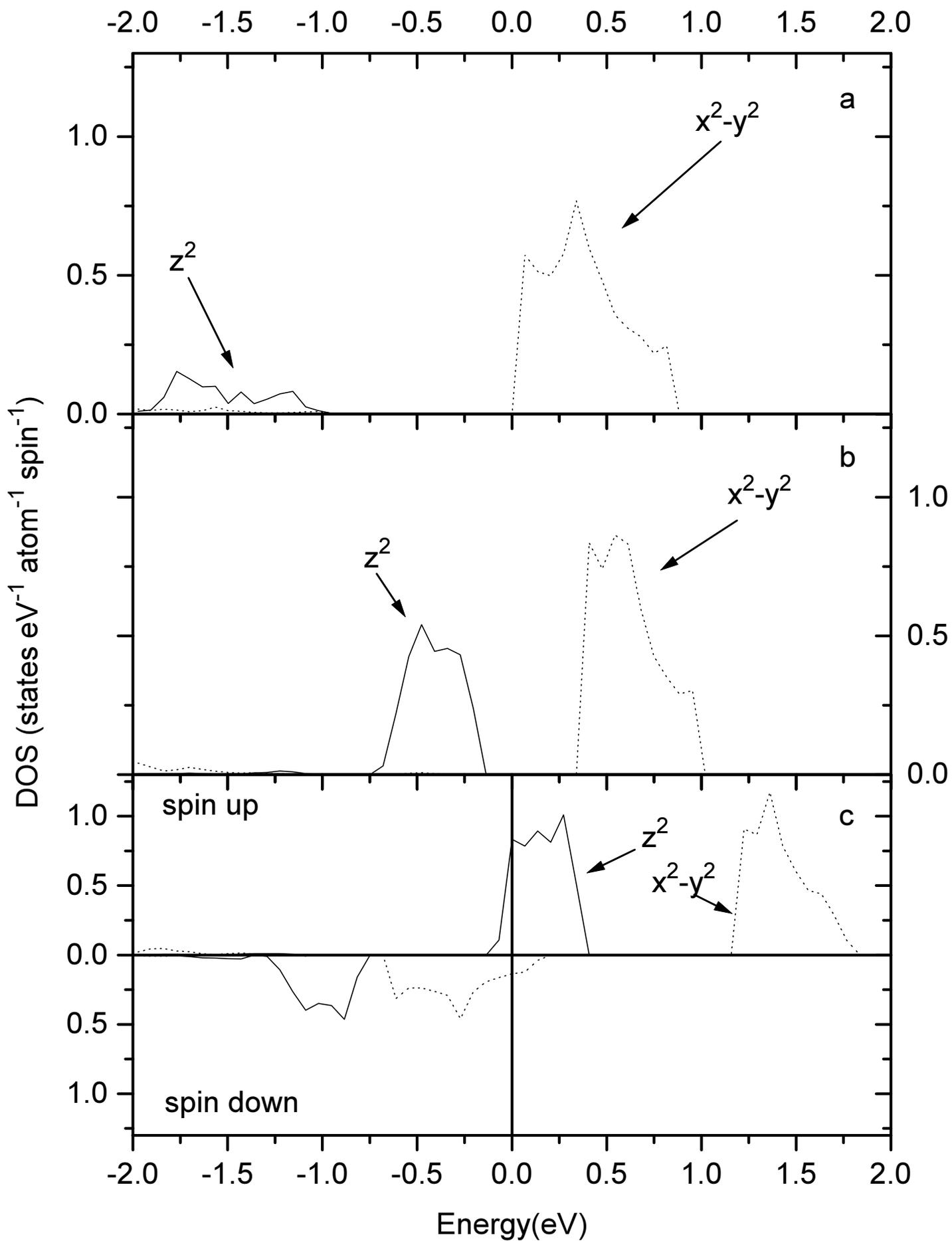
\end{document}